\documentclass[aps,prl,twocolumn,showpacs,amsmath,amssymb,superscriptaddress,floatfix]{revtex4-1}

\usepackage{graphicx}
\usepackage{dcolumn}
\usepackage{bm}
\usepackage{amssymb}
\usepackage{booktabs}
\usepackage{color}
\usepackage{microtype}
\usepackage[T1]{fontenc}

\usepackage[papersize={8.5in,11in}]{geometry}
\usepackage[colorlinks,plainpages=false,linkcolor=blue,urlcolor=blue,citecolor=blue,pdfpagemode=UseNone,pdfstartview=FitBH]{hyperref}

\usepackage{sidecap}
\renewcommand{\figurename}{Fig.}
\renewcommand{\tablename}{Table}
\makeatletter\renewcommand{\fnum@figure}[1]{\figurename~\thefigure~(color online).}\makeatother
\makeatletter\renewcommand{\fnum@table}[1]{\tablename~\thetable.}\makeatother

\newcommand{\beq}{\begin{equation}}
\newcommand{\eeq}{\end{equation}}

\newcommand{\RNO}{$R$NiO$_3$}
\newcommand{\mi}{\mathrm{i}}

\begin{document}

\title{Origins of bond and spin order in rare-earth nickelate bulk and heterostructures}

\author{Yi~Lu}
\affiliation{Max-Planck-Institut f\"ur Festk\"orperforschung, Heisenbergstrasse~1, 70569 Stuttgart, Germany}

\author{Zhicheng~Zhong}
\affiliation{Max-Planck-Institut f\"ur Festk\"orperforschung, Heisenbergstrasse~1, 70569 Stuttgart, Germany}

\author{Maurits~W.~Haverkort}
\affiliation{\mbox{Max-Planck-Institut f\"ur Chemische Physik fester Stoffe, N\"othnitzer Strasse 40, 01187 Dresden, Germany}}
\affiliation{\mbox{Institut f\"ur Theoretische Physik, Ruprecht-Karls-Universit\"at Heidelberg, Philosophenweg 19, 69120 Heidelberg, Germany}}

\author{Philipp~Hansmann}
\affiliation{Max-Planck-Institut f\"ur Festk\"orperforschung, Heisenbergstrasse~1, 70569 Stuttgart, Germany}
\affiliation{Institut f\"ur Theoretische Physik, Eberhard Karls Universit\"at T\"ubingen, Auf der Morgenstelle 14, 72076 T\"ubingen}

\date{\today}

\begin{abstract}
We analyze the charge- and spin response functions of rare-earth nickelates \RNO{} and their heterostructures using random-phase approximation in a two-band Hubbard model. The inter-orbital charge fluctuation is found to be the driving mechanism for the rock-salt type bond order in bulk \RNO, and good agreement of the ordering temperature with experimental values is achieved for all \RNO{} using realistic crystal structures and interaction parameters. We further show that magnetic ordering in bulk is not driven by the spin fluctuation and should be instead explained as ordering of localized moments. This picture changes for low-dimensional heterostructures, where the charge fluctuation is suppressed and overtaken by the enhanced spin instability, which results in a spin-density-wave ground state observed in recent experiments. Predictions for spectroscopy allow for further experimental testing of our claims.
\end{abstract}

\maketitle

\textit{Introduction.---}
Understanding the mechanisms behind collective orders and excitations in solids is a pivotal topic in current condensed-matter research. The interplay between various electronic degrees of freedom at different time and energy scales gives rise to virtually unlimited variety of properties such as metal-insulator transitions (MIT), multiferroicity and superconductivity. One example of long-standing interest are the rare-earth nickelates \RNO{}, which exhibit complex ordering phenomena depending on the NiO$_6$ octahedra tilts and distortions controlled by the radius of rare-earth ion $R$~\cite{Torrance1992,Medarde1997,Catalan2008}. For the smallest $R=$ Lu, \RNO{} goes through a MIT at $T_c\simeq$ 600~K, accompanied by a rock-salt type bond order of NiO$_6$ octahedra at wave vector $\bm{q}_c = (1/2,1/2,1/2)$ (in units of $2\pi/a$ with $a$ the pseudocubic lattice constant) with alternating Ni-O bond lengths. An antiferromagnetically ordered phase follows at much lower temperature $T_s\simeq$ 130~K with an unusual $\bm{q}_s = (1/4, 1/4, 1/4)$. The temperature difference between the two transitions decreases with increasing $R$ size and disappears at $R$ = Nd with $T_c=T_s\simeq$ 200~K. LaNiO$_3$, with the largest $R$, remains metallic at all temperatures. This complex phase diagram can be further enriched by newly developed controlled growth of oxides with atomic precision~\cite{Hwang2012}. Recent experiments have shown that via strain, dimensionality, and symmetry control in epitaxial films and heterostructures, the phase boundaries can be shifted and different order parameters can be selectively altered~\cite{Scherwitzl2011,Liu2011,Boris2011,Frano2013,Lu2016,Hepting2014,Catalano2015,Hoffman2013,Meyers2016,Kim2016}. The quasi-two-dimensional heterostructures, for instance, show a pure spin-density-wave (SDW) ground state without bond order~\cite{Frano2013,Lu2016,Hepting2014}---remarkably different from the bulk.

The complex phase behavior of the nickelates and the apparent dichotomy between the bulk and heterostructures pose several theoretical challenges archetypical for transition metal oxides. The outstanding challenge is to understand the relation between the structural and electronic transitions. Recent discussions in the context of ``negative charge transfer'' insulators~\cite{Mizokawa1991} have shown that the bond order is indispensable for understanding the MIT of the \RNO{}. Constraining the system to the experimentally observed bond-ordered state, an insulating ground state was found in small-cluster~\cite{Johnston2014,Green2016}, mean-field~\cite{Lau2013,Johnston2014}, and dynamical mean-field~\cite{Park2012,Subedi2015} calculations. However, the origin of the essential bond order, or its absence in low-dimensional heterostructures, has remained obscure.

In this Letter, we address this crucial issue by examining---on equal footing---the charge- and spin response functions in the unordered metallic phase for the \RNO{} series with multiorbital random phase approximation (RPA)~\cite{Takimoto2004,Graser2009} in a two-band Hubbard model. We identify a dominating charge response at $\bm{q}_c$ originated from inter-orbital fluctuations in the Ni-$e_g$ states, which can drive the system into the bond order via strong electron-phonon coupling~\cite{Medarde1998}. The instability increases with increasing $Pbnm$ (or $R\bar3c$ for $R$ = La) distortion and naturally explains the $R$ dependence of the ordering temperature $T_c$ in bulk \RNO{}. The previously assumed primary spin instability~\cite{Lee2011a,Lee2011b}, on the other hand, remains marginal in \emph{all} bulk \RNO. We further show that charge fluctuations are suppressed in spatially confined heterostructures below certain thickness, and a concomitant increase in the spin response can give rise to the experimentally observed SDW ground state without bond order~\cite{Frano2013,Lu2016,Hepting2014}.
% END INTRO

% METHOD PART
\textit{The Hamiltonian and multiorbital RPA.---}
We consider an effective two-band model~\cite{Kubo2007} for the Ni-$e_g$ orbitals
\begin{align}\label{eq:ham}
\begin{split}
H = 
& \sum_{\bm{k}\sigma ab}\epsilon_{\bm{k}ab} c^\dag_{\bm{k}a\sigma} c_{\bm{k}b\sigma} + U\sum_{ia} n_{ia\uparrow}n_{ia\downarrow}+U^\prime\sum_{i}n_{ia}n_{ib}\\
& + J \sum_{i\sigma\sigma^\prime}c^\dag_{ia\sigma}c^\dag_{ib\sigma'}c_{ia\sigma'}c_{ib\sigma}  + J^\prime \sum_{i}c^\dag_{ia\uparrow}c^\dag_{ia\downarrow}c_{ib\downarrow}c_{ib\uparrow},
\end{split}
\end{align}
where $c^\dag_{ia\sigma}$ ($c^\dag_{\bm{k}a\sigma}$) creates an electron at site $i$ (momentum $\bm{k}$) in orbital $a$ with spin $\sigma = \uparrow, \downarrow$. The orbital indices $a,b \in \{d_{3z^2-1},d_{x^2-y^2}\}$ label the $e_g$ Wannier functions. $\epsilon_{\bm{k}ab}$ is the hopping matrix including the chemical potential. The number operators $n_{ia\sigma}=c^\dag_{ia\sigma}c_{ia\sigma}$ and $n_{ia} = n_{ia\uparrow} + n_{ia\downarrow}$. The coupling constants $U$, $U'$ denote the strength of intra- and inter-orbital Coulomb repulsion, and $J$, $J'$ the intraorbital exchange and pair hopping. The RPA charge and spin susceptibilities are then given as
\beq\label{eq:rpa}
\hat \chi^{c/s}=\hat \chi^0(\mathbb{I}\pm\hat \chi^0 U^{c/s})^{-1}
\eeq
where the matrix elements of the bare susceptibility $\hat \chi^0$ reads
\beq\label{eq:chi0}
\chi^0_{aa'bb'}(\bm{q},\mi \Omega_n) = -\frac{1}{\beta} \sum_{\bm{k} m} G^0_{ab'}(\bm{k},\mi \omega_m) G^0_{ba'}(\bm{k}+\bm{q},\mi \omega_m')
\eeq
with $\beta=1/T$ the inverse temperature and $G^0_{ab'}$ the bare Green's function. $\omega_m$ and $\omega_m'=\omega_m+\Omega_n$ are the fermionic Matsubara frequencies. $U^c$ and $U^s$ are the the bare vertices coupling to charge- and spin-type of fluctuations, respectively, with matrix elements $U^c_{aa'bb'}=(U,-U'+2J,2U'-J,J',0)$ and $U^s_{aa'bb'}=(U,U',J,J',0)$ when ($a = a' = b = b'$, $a = b' \neq a' = b$, $a=a'\neq b = b'$, $a=b\neq a' = b'$ and otherwise). The total charge/spin susceptibility is then $\chi^{c/s}=\frac{1}{2}\sum_{ab}[\chi^{c/s}_{abba}]$.

While the interaction constants are often adopted as tuning parameters~\cite{Takimoto2004,Graser2009}, it would be favorable to take parameters most relevant to the specific materials at hand. Such effective parameters can be calculated from first principles using the constrained RPA~\cite{cRPA}. For LuNiO$_3$ the values in the $e_g$ subspace are calculated by Seth \emph{et al.}~\cite{Seth2016} as $U=1.65$~eV, $J=0.33$~eV, $U'=U-2J$ and $J'=J$. These values are considerably smaller than the typical \RNO{} bandwidth of $\sim$3~eV~\cite{SupMat}---a parameter regime that RPA is well suited for. It is, however, important to note that RPA ignores crucial vertex corrections and overestimates the instabilities when using bare interaction parameters. Therefore we use the renormalized values given by the particle-particle vertex equation $\hat{\bar U} = \hat U (\mathbb{I}+\hat U \hat \Gamma^{p})^{-1}$ with $\Gamma^{p}_{aa'bb'}(\bm{q},\mi \Omega_n)=\frac{1}{\beta} \sum_{\bm{k} m} G^0_{ab'}(\bm{k},\mi \omega_m) G^0_{ba'}(-\bm{k}+\bm{q},\mi \omega_m')$. Such an approach has shown to reproduce correctly the exact susceptibilities obtained by quantum Monte Carlo methods in Hubbard models~\cite{Chen1991,Bulut1993}. We arrive at static renormalized values at $T=300$~K with $\bar U$ = 1.02~eV, $\bar U'$ = 0.70~eV, $\bar J$ = 0.17~eV and $\bar J'$ = 0.13~eV by averaging over the \RNO{} series. While the exact values of these parameters have a certain material and temperature dependence, we have checked that the variation does not change the results substantially. For simplicity we keep the interaction parameters fixed throughout this Letter unless otherwise noted.
% END OF METHOD PART

% START DISCUSSION PART FIG1
\begin{figure}[t]
  \includegraphics[width=\columnwidth]{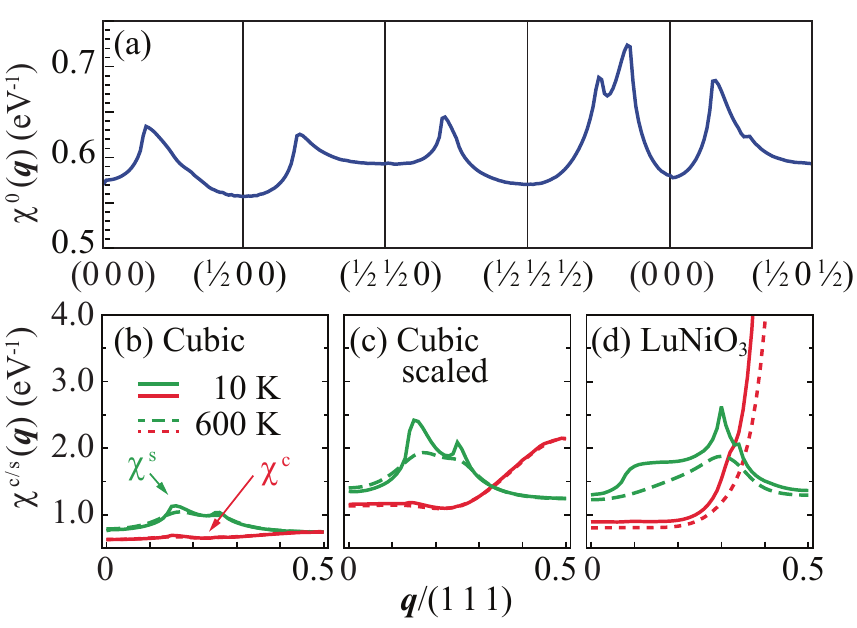}
  \caption{\label{fig:chi} (a) Static $\chi^0(\bm{q})$ along the high symmetry lines for cubic LaNiO$_3$ at $T$ = 10~K. (b)-(d) Temperature dependence of $\chi^c(\bm{q})$ (red) and $\chi^s(\bm{q})$ (green) for (b) cubic LaNiO$_3$ with original and (c) scaled bandwidth (see text), and (d) LuNiO$_3$.}
\end{figure}

To study the structural dependence of charge- and spin response functions, we performed calculations for the experimentally determined \RNO{} structures in the $Pbnm$ or $R\bar3c$ metallic phase~\cite{Alonso2001}. A hypothetical cubic LaNiO$_3$~\cite{SupMat} was also included as a reference system. The hopping matrices $\epsilon_{\bm{k}ab}$ are constructed using maximally localized Wannier orbitals~\cite{wan90} obtained from density functional (DFT) calculations~\cite{wien2k}. To formulate the RPA calculation within a two-band model, the bare Green's functions are unfolded~\cite{Ku2010} to the pseudocubic Brillouin zone (BZ) for the non-cubic cases. 

We start by discussing the static non-interacting $\chi^0(\bm{q})$ for LaNiO$_3$ constrained to cubic symmetry shown in Fig.~\ref{fig:chi}(a). In agreement with previous results~\cite{Lee2011a,Lee2011b}, two maxima are found at incommensurate wave vectors around $\bm{q}_s$. Their transition to the commensurate $\bm{q}_s$ can occur due to the spin-lattice coupling or simply by adopting hopping parameters better describing the experimentally measured Fermi surface~\cite{Lee2011a}, and therefore we refer to them as $\bm{q}_s$ hereafter. In a single-band RPA description, the spin instability is expected to be dominating with repulsive $U$ as $\chi^0$ is positive and $\chi^s$ scales with $(1-\chi^0U)^{-1}$, while the charge response $\chi^c\propto(1+\chi^0U)^{-1}$ is always suppressed. In the multiorbital case, however, this simple argument does not hold due to the matrix nature of Eq.~\eqref{eq:rpa}. An increase of the charge response at $\bm{q}_c$---which corresponds to a minimum of $\chi^0(\bm{q}$)---appears, once the inter-orbital interaction $U'$ is included~\cite{SupMat}. Based on the observations that both $\chi^c(\bm{q})$ and $\chi^s(\bm{q})$ show instabilities at the respective experimental wave vectors for bond- and magnetic order, one naturally poses the question if the dependence of $\chi^s$ and $\chi^c$ upon the $Pbnm$ (or $R\bar3c$ for unconstrained LaNiO$_3$) distortion can explain the material trend of phase transitions in the \RNO{} series.

The distortions affect the material dependent $\epsilon_{\bm{k}ab}$ in two distinctive aspects: i) an overall reduction of bandwidth, and ii) broken ``selection rules'' for orbital transitions due to lower symmetry. The effect of i) on the response functions is shown in Fig.~\ref{fig:chi}(b) and (c). Both $\chi^s(\bm{q})$ and $\chi^c(\bm{q})$ show a noticeable increase when the bandwidth of the cubic LaNiO$_3$ ($\approx 3.9$eV) is reduced to that of orthorhombic LuNiO$_3$ ($\approx 2.7$eV)~\cite{SupMat}. Subsequently, we see the effect of ii) when comparing the scaled cubic case to the actual calculation of LuNiO$_3$ shown in Fig.~\ref{fig:chi}(d). While $\chi^s(\bm{q})$ changes slightly its momentum dependence without noteworthy increase of the overall response, $\chi^c(\bm{q})$ becomes dominant and approaches divergence at $\bm{q}_c$ below 600~K, which signals a phase transition to an ordered state with ordering vector $\bm{q}_c$, in agreement with experiment.

\begin{figure}[t]
  \includegraphics[width=\columnwidth]{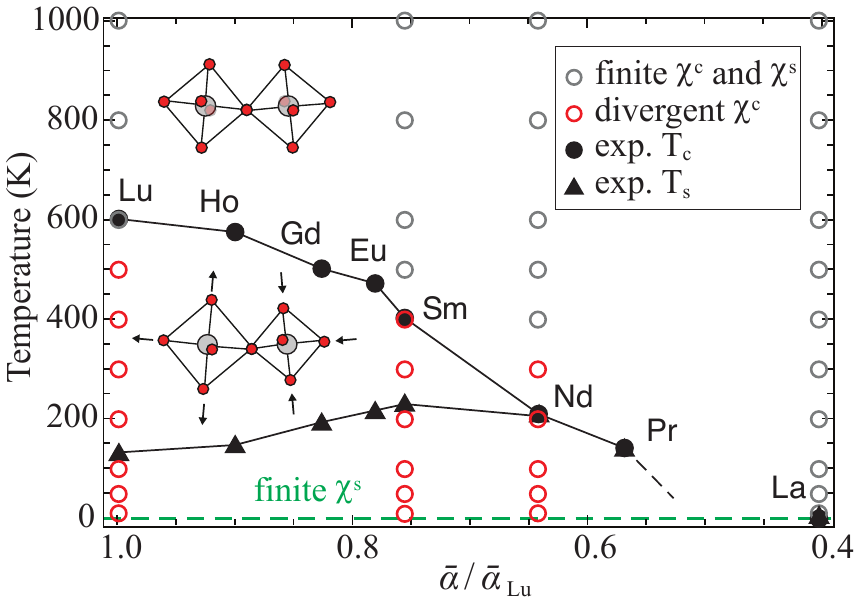}
  \caption{\label{fig:pd} Phase diagram of \RNO{} with different distortions. $\bar \alpha$ is the averaged deviation of the Ni-O-Ni bond angle from 180$^\circ$. The open circles mark the calculated data points, where red color indicates the divergence of $\chi^c(\bm{q}_c)$. $\chi^s(\bm{q}_s)$ remains finite at all temperatures, which is represented by a dashed line at $T$= 0. The experimental $T_c$ ($T_s$) values~\cite{Catalan2008} are denoted by black dots (triangles).}
\end{figure}

% START DISCUSSION PART FIG2
\textit{The bulk phase diagram.---}
Figure~\ref{fig:pd} shows an overview of the calculated temperature dependence of $\chi^c(\bm{q}_c)$ and $\chi^s(\bm{q}_s)$ for \RNO{} with various distortions in addition to the extremal case of LuNiO$_3$. The charge response function $\chi^c(\bm{q}_c)$ dominates over the whole \RNO{} series and the boundary of its divergence follows closely the experimental $T_c$, including the absence of a divergence/transition for LaNiO$_3$ down to the lowest considered temperature. The spin response function $\chi^s(\bm{q}_s)$, on the other hand, remains finite for all materials throughout the full considered temperature range down to 10~K, which indicates a secondary role of spin fluctuations. 
We emphasize that this is also true for NdNiO$_3$---with experimentally equal $T_c$ and $T_s$---which first and foremost undergoes a charge driven transition. A direct consequence is that the subsequent magnetic transition should be understood starting from the insulating bond-ordered state (a more apparent statement for compounds with smaller $R$). Hence, instead of an itinerant approach based on Fermi liquid~\cite{Lee2011a,Lee2011b}, the magnetic order in \RNO{} may be more appropriately studied using a localized spin model. Another fact supporting this claim is the distortion dependence of the experimental $T_s$ that increases with decreasing structural distortion (or increasing bandwidth), while an opposite trend should be expected if it is driven by $\chi^s(\bm{q}_s)$. In fact, in the part of the phase diagram where transition to the bond ordered state and magnetic transition are separated, $T_s$ is proportional to the exchange interaction $J_{ex}\sim W^2/\Delta$ given by perturbation theory, with $W$ the bandwidth and $\Delta$ the characteristic charge excitation gap in the insulating phase defined by the Coulomb interaction and charge transfer energy. This is also confirmed by the energy gain of the antiferromagnetic state which we calculated with constrained DFT+$U$~\cite{SupMat} in the low-temperature monoclinic phase. The peculiar out-of-trend behavior of $T_s$ for $R$ = Nd and Pr is naturally explained since the magnetic order can only occur in the insulating bond-ordered state. This also explains the elevated $T_s$ in a NdNiO$_3$ film when $T_c$ is increased by epitaxial strain~\cite{Catalano2015}.

\begin{figure}[t]
  \includegraphics[width=\columnwidth]{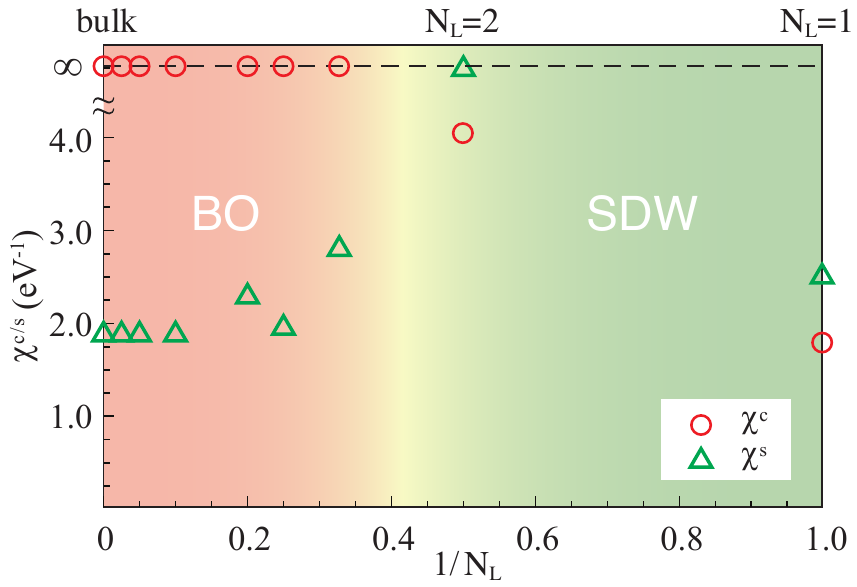}
  \caption{\label{fig:confine} Thickness $N_L$ dependence of $\chi^c(\bm{q}_c)$ and $\chi^s(\bm{q}_s)$ at $T$ = 10~K for NdNiO$_3$ with $N_L$ = 1--5, 10, 20 and 40. The bulk ($1/N_L$= 0) values are plotted for comparison.}
\end{figure}

%START DISCUSSION OF CONFINEMENT
\textit{Effect of spatial confinement.---} After providing a unified description for the transitions in bulk \RNO{} by identifying a dominating-bond-order picture, we are left with a puzzle as to how this is applicable to low-dimensional heterostructures, where magnetic order is observed without bond order~\cite{Frano2013,Lu2016,Hepting2014}. To understand the dimensional effect we performed calculations for NdNiO$_3$ slabs with different thicknesses of $N_L$ layers~\cite{SupMat}. The lattice symmetry and hopping parameters were kept the same as the bulk, leaving the dimensionality as the only control parameter. The thickness dependence of $\chi^c(\bm{q}_c)$ and $\chi^s(\bm{q}_s)$ is shown in Fig.~\ref{fig:confine}. For $N_L\geq$ 10, the details of $\chi^{c/s}(\bm{q})$ remain largely unaffected compared to the bulk~\cite{SupMat} with their respective maxima at $\bm{q}_c$ and $\bm{q}_s$ closely reproducing the bulk values. For $N_L$ below 5, deviations from the bulk are noticeable in the details~\cite{SupMat}, and a dimensional crossover can be observed between $N_L=$ 3 and 2, where $\chi^{c}(\bm{q}_c)$ is suppressed while $\chi^{s}(\bm{q}_s)$ becomes dominant and even diverges with $N_L$ = 2. For $N_L$ = 1, the system becomes two-dimensional with $\chi^{c}(\bm{q}^\parallel_c)$ ($\,\parallel$ denotes the $\bm{q}$ projection in the layer plane) fully suppressed~\cite{SupMat}. Interestingly, while $\chi^{s}(\bm{q}^\parallel_s)$ remains dominating in single layer as in the bilayer, it does not diverge~\footnote{Note that in principle larger interaction parameters should be used for the two-dimensional case due to the less effective screening, which may still push $\chi^{s}(\bm{q}_s)$ and/or $\chi^{c}(\bm{q}_c)$ to divergence.}. The difference might be attributed to a better nesting condition in the bilayer~\cite{SupMat}. The seemingly contradicting observations in heterostructures are thus explained by the suppression of $\chi^{c}(\bm{q}_c)$ and enhancement of $\chi^{s}(\bm{q}_s)$ in reduced dimensions, although we note that the exact critical thickness $N_L$ may differ for, e.g., different materials and/or epitaxial strains. These findings further prove the validity of our analysis and in addition point out an itinerant origin of the magnetism in heterostructures, qualitatively different from the bulk materials.

\begin{figure}[t]
  \includegraphics[width=\columnwidth]{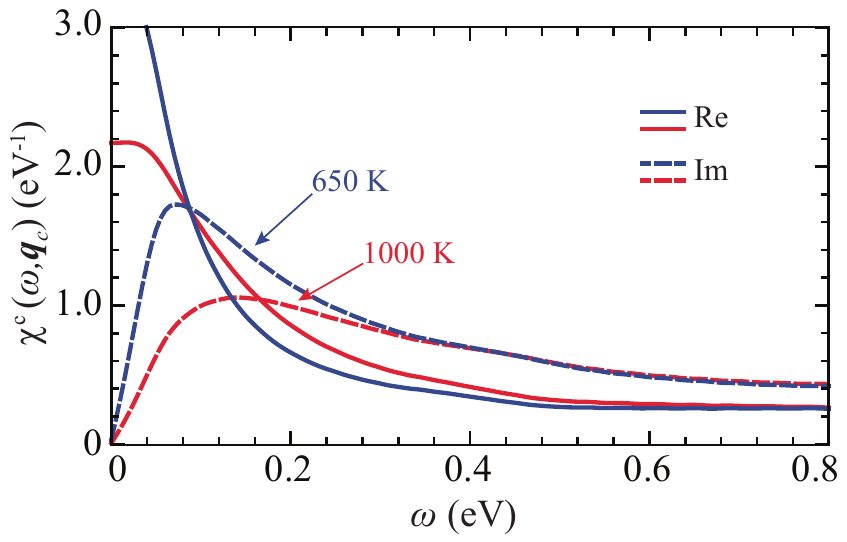}
  \caption{\label{fig:dyn} Real (solid) and imaginary (dashed) parts of $\chi^c(\omega,\bm{q}_c)$ for LuNiO$_3$ at 1000~K (red) and 650~K (blue).}
\end{figure}

% START DISCUSSION OF DYNAMIC RESPONSE
\textit{Dynamics of the charge response.---}
The divergence of static susceptibilities yields information about critical parameters and symmetry of the phase transition to an ordered state. The frequency dependence of $\chi^{\mathrm{c/s}}(\omega,\bm{q})$, on the other hand, provides additional information about the evolution of the characteristic correlation time of a charge or spin fluctuation when approaching the phase transition. Such quantities, when experimentally accessible, can further strengthen or falsify our proposals. Figure~\ref{fig:dyn} shows the real and imaginary part of $\chi^{\mathrm{c}}(\omega,\mathbf{q}_c)$ for LuNiO$_3$ at two different temperatures above $T_c$. With decreasing temperature, the spectral weight of the imaginary part shifts to lower frequencies with a concomitant increase of the Kramers-Kronig related real part at $\omega=0$. At 650 K, a temperature close to $T_c \simeq 600$~K, the maximum of the peak is at $0.075$~eV corresponding to a timescale of $\sim 10^{-13}$~s. This timescale is slower than that of core-level or optical spectroscopy on the order of a few to dozen femtoseconds and should be detectable using such methods. Indeed, signs of dynamic valence fluctuation of Ni associated to the bond order was observed in LuNiO$_3$ above $T_c$ using x-ray absorption~\cite{Medarde2009}. The charge fluctuation was also indicated by the Fermi surface reconstruction with $\bm{q}_c$ observed in metallic LaNiO$_3$ films using angle-resolved photoemission~\cite{Yoo2015}. Future resonant inelastic x-ray scattering (RIXS) experiments, which can measure the response functions directly in the frequency domain, may help to gain more quantitative information about the dynamics of the fluctuations and its relation to the phase transition.

\textit{Closing remarks.---}
In the following we discuss briefly the order of the transitions. Experimentally, the MIT in bulk \RNO{} is weakly first order, and the magnetic transition is second order when separated from the first transition. Although the divergence of $\chi^c(\bm{q})$ is always an indication of a second-order transition, the first-order nature of the transition at $T_c$ is already accounted for by considering its coupling to the lattice~\cite{Medarde1998} and the resulting bond order. It would be interesting to study the order of isostructural MIT in \RNO{} that has been possibly realized in epitaxial films~\cite{Catalano2015,Meyers2016} by restraining the lattice symmetry to structurally inert substrates. The lack of hysteresis across the MIT in the dc transport data for one film in Ref.~\cite{Catalano2015} may already be indicative of a second-order transition. In addition, RIXS measurement as mentioned above can also give insights to the order of the transition. A divergence of spectral weight and collapse in energy at $\bm{q}_c$ should occur for a second-order transition when approaching the transition temperature from the metallic phase. For the heterostructures, the SDW transition was revealed to be second order~\cite{Hepting2014}, in agreement with our study.

A second remark is on the relation of our work to earlier approaches starting from the local limit which include the O-$p$ orbitals explicitly~\cite{Johnston2014,Green2016,Lau2013,Johnston2014,Park2012}. In the negative charge transfer picture, the oxygens donate one electron onto each Ni site in the metallic state. Upon bond order, the oxygen holes condensate onto half of the NiO$_6$ octahedra, which in the extreme case gives rise to a local $S=0$ singlet state while the other half of the Ni sublattice develops into a Mott phase with $S=1$. The traditionally termed ``charge order'' in the bond-ordered phase therefore does not involve actual charge redistribution on the Ni sites. Our current findings do not contradict this local many-body picture. The Ni $e_g$ Wannier functions are composite objects including hybridized Ni-$d$ states and the neighboring O-$p$ states. While the divergence of $\chi^c(\bm{q}_c)$ in our calculation indicates a nearest neighbor rock-salt type ``charge order'', the extended tails of the Wannier functions on the O sites ensure that large part of the charge density does not move in space.

\textit{Conclusion.---}
We have presented a study of charge and spin response functions for the family of rare-earth nickelates \RNO. Within multiorbital RPA approach on an effective Ni-$e_g$ model, we showed that the inter-orbital fluctuation increases with $Pbnm$ distortion and strongly contributes to the charge response function $\chi^c(\bm{q})$ that is responsible for the observed bond order in bulk \RNO. The charge instability is suppressed in low-dimensional heterostructures, leaving magnetism to prevail, in agreement with recent experimental observations. The frequency dependence of the calculated charge response explains the dynamic charge fluctuation above the phase transition observed in x-ray absorption and photoemission experiments, and stands as a prediction for future experiments.

We thank B. Keimer, E. Benckiser, X. Cao, M. H\"oppner, G. Khaliullin, and O.~K. Andersen for motivation and fruitful discussions. We are especially grateful
to authors of Ref.~\cite{Seth2016} for communicating us with cRPA results before publication.

\clearpage

\appendix 
\onecolumngrid

\begin{center}
\textbf{\large{Supplemental Material}}
\end{center}

\setcounter{figure}{0}
\setcounter{table}{0}
\setcounter{equation}{0}
\renewcommand{\thefigure}{S\arabic{figure}}
\renewcommand{\thetable}{S\Roman{table}}
\renewcommand{\theequation}{S\arabic{equation}}

\section{Interorbital fluctuations and divergence of $\chi^c(\mathbf{q})$}

\begin{figure}[htb]
  \includegraphics[width=0.5\textwidth]{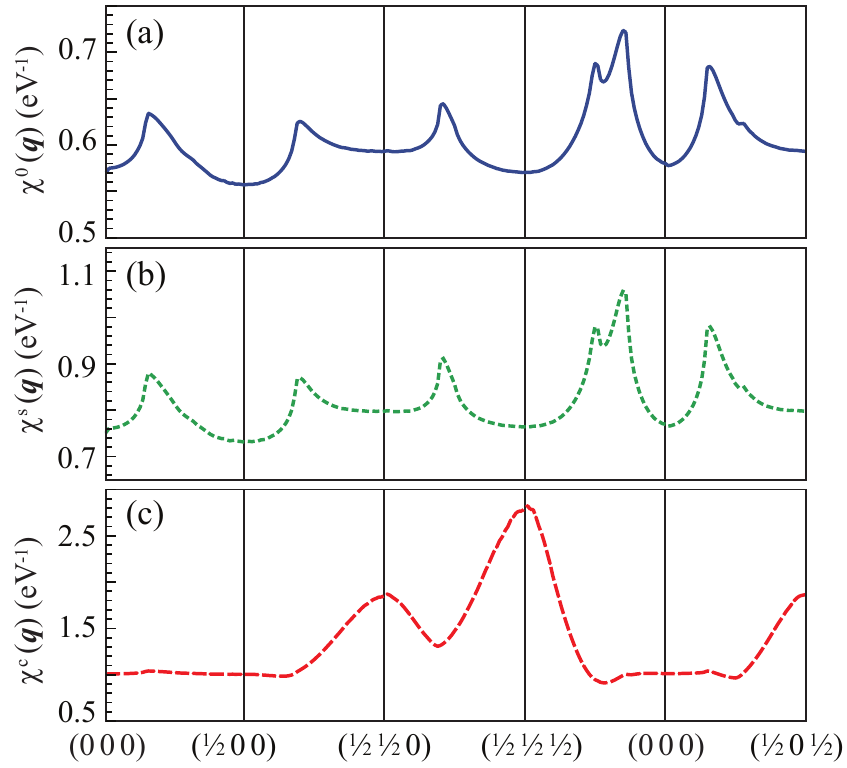}
  \caption{\label{Sfig:cubic_chi} Static (a) $\chi^0(\bm{q})$, (b) $\chi^s(\bm{q})$ and (c) $\chi^c(\bm{q})$ along the high symmetry lines in the cubic BZ with $U=U'=$ 0.85~eV and $J=J'=$ 0.}
\end{figure}

To illustrate the multi-orbital origin of the divergence of the static charge $\chi^c$, we analyze the response functions in a simplified scenario using a cubic LaNiO$_3$ structure with lattice constants optimized in DFT. Fig. 1(a) is reproduced in Fig.~\ref{Sfig:cubic_chi}(a) for the ease of discussion.

For simplicity, we consider the plane with $q_x=q_y$ where both $\bm{q}_c$ and $\bm{q}_s$ reside. The non-zero entries of the bare susceptibility $\chi^0_{aa'bb'}$ have indices $aaaa$, $aabb$, $abba$, $abab$ with parity between $a,b=$ 1 ($d_{3z^2-1}$), 2 ($d_{x^2-y^2}$) or 2, 1 due to the cubic symmetry. The diverging element for the spin channel is then $\chi^s_{1111}$ with the critical interaction given as $(U-J')^s_{\mathrm{crit}} = 1/\mathrm{max}[(\chi^0_{1111} + \chi^0_{1212})]$, while for the charge channel $\chi^s_{1221}$ with $(U'-J)^c_{\mathrm{crit}} = 1/\mathrm{max}[3(\chi^0_{1221} - \chi^0_{1122})]$. One immediately notices that by neglecting the small pair hopping term $\chi^0_{1212}$, the single-band RPA result for the spin channel is recovered with the divergence determined by the intraorbital fluctuation $\chi^0_{1111}$. On the other hand, the interorbital fluctuation $\chi^0_{1221}$ may give rise to a comparable or even stronger charge instability given that $\mathrm{max}[\chi^0_{1221}]\gtrsim\frac{1}{3}\mathrm{max}[\chi^0_{1111}]$. Such an effect is shown in Fig.~\ref{Sfig:cubic_chi}(b) and (c) which plot the static RPA $\chi^s$ and $\chi^c$ with $U=U'=0.85$~eV and $J=J'=0$. Note that the parameters chosen here are only for demonstration purposes. While $\chi^s$ is approximately a scaled-up version of $\chi^0$ ---reminiscent to the single-band RPA result---with a leading instability at $\bm{q}_s$, $\chi^c$ develops a strong instability at the experimental bond-order wave vector $\bm{q}_c$.

\section{The $e_g$ band structure of bulk \emph{R}NOs}

\begin{figure*}[htb]
  \includegraphics[width=\textwidth]{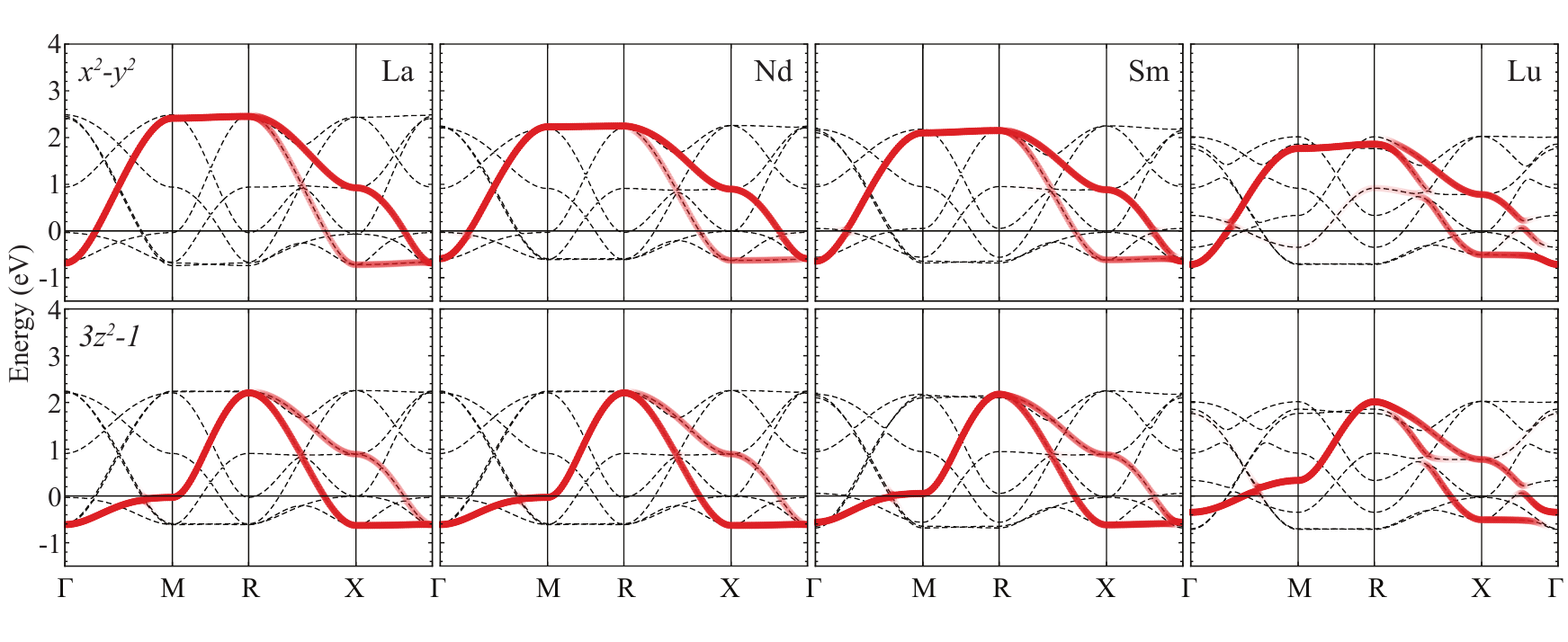}
  \caption{\label{Sfig:bands} The $e_g$ band structures of the \RNO{} series along the high-symmetry lines $\Gamma (000)$-M$(\frac{1}{2}\frac{1}{2}0)$-R$(\frac{1}{2}\frac{1}{2}\frac{1}{2})$-X$(\frac{1}{2}00)$-$\Gamma (000)$ in the pseudocubic BZ. The rare-earth $R$=La, Nd, Sm, and Lu. The unfolded $x^2-y^2$ and $3z^2-1$ characters are highlighted red in the upper and lower panels, respectively.}
\end{figure*}

Figure~\ref{Sfig:bands} shows the $e_g$ band structures for \RNO s with $R$=La, Nd, Sm, and Lu. With increasing structural distortion, the bandwidth reduces from 3.1~eV for La to 2.7~eV for Lu. An increased intermixing between $x^2-y^2$ and $3z^2-1$ characters can also be observed, most noticeably along R-X-$\Gamma$.

\section{Overview of response functions for bulk \emph{R}NOs}

Comparison of the charge- and spin response functions at two selected temperatures $T=$ 10 and 600~K for \RNO{} with $R=$ Lu, Sm, Nd and La. The charge instability at $\bm{q_c}$ dominates over the spin one around $\bm{q_s}$ in all compounds.

\begin{figure*}[htb]
  \includegraphics[width=0.5\textwidth]{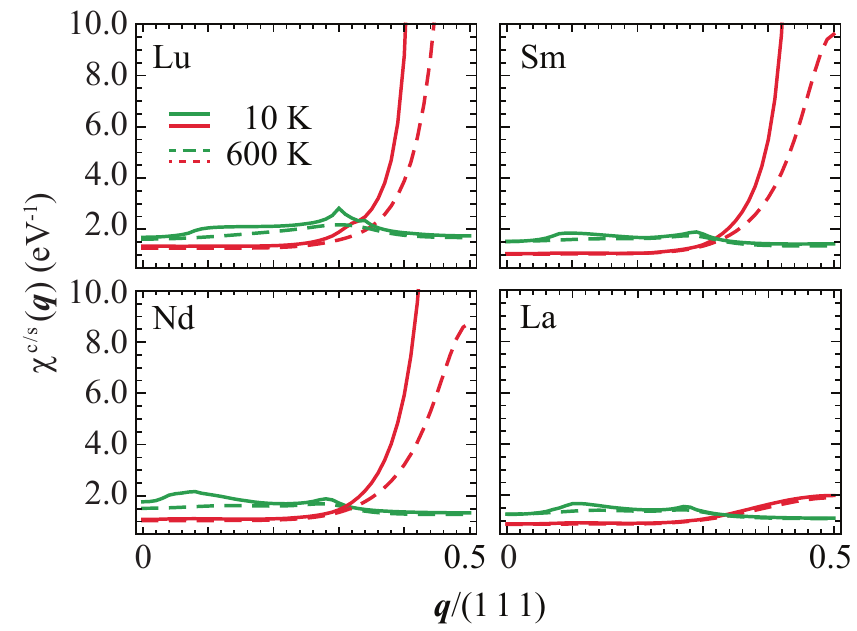}
  \caption{\label{Sfig:chiall} Temperature dependence of $\chi^s(\bm{q})$ (green) and $\chi^c(\bm{q})$ (red) for different \RNO. }
\end{figure*}

\section{Doping- and crystal-field dependence of response functions}

The charge- and spin response functions for \RNO{} can be further affected by e.g., doping and crystal-field splitting of the $e_g$ states. The doping can either be introduced chemically or via charge transfer through transition-metal-oxide interfaces [S1]. The crystal-field splitting is common in strained epitaxial thin films. Here we study these effects using NdNiO$_3$ as an example. 

\subsection{a. The effect of doping}

\begin{figure*}[htb]
  \includegraphics[width=0.5\textwidth]{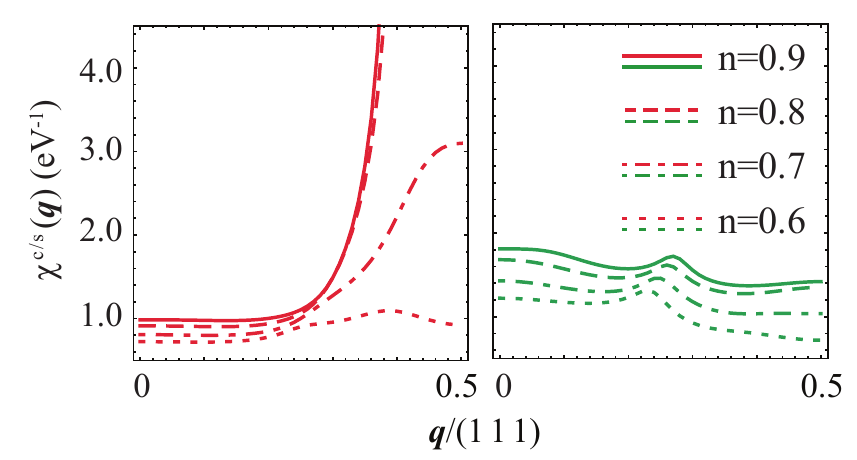}
  \caption{\label{Sfig:chihole} Hole doping dependence of $\chi^c(\bm{q})$ (red) and $\chi^s(\bm{q})$ (green) for NdNiO$_3$ at 300 K.}
\end{figure*}

\begin{figure*}[htb]
  \includegraphics[width=0.5\textwidth]{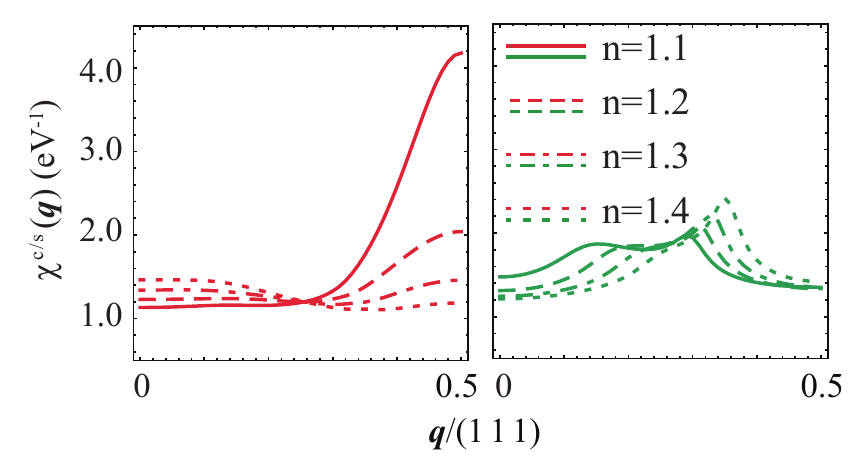}
  \caption{\label{Sfig:chielec} Electron doping dependence of $\chi^c(\bm{q})$ (red) and $\chi^s(\bm{q})$ (green) for NdNiO$_3$ at 300 K.}
\end{figure*}

Figure~\ref{Sfig:chihole} and \ref{Sfig:chielec} show the doping dependence of the charge and spin response functions for hole and electron doping, respectively. The charge instability is relatively robust against hole doping and is only significantly reduced for doping above 30\%. On the other hand, $\chi^s(\bm{q})$ only shows a small change in $\bm{q}$ dependence due to the change of Fermi surface and a gradual overall decrease. 

Compared to the hole doping, electron doping is more efficient in removing the charge instability. Interestingly, the spin instability is seen to increase with doping, with a concomitant shift of the maximum towards $(1/2,1/2,1/2)$. This suggests doping can serve as a promising route for controlling the magnetism in the \RNO{} and may shed light to the nontrivial magnetic properties in some \RNO{} heterostructures [S2].

\subsection{b. The effect of crystal field}

\begin{figure*}[htb]
  \includegraphics[width=0.5\textwidth]{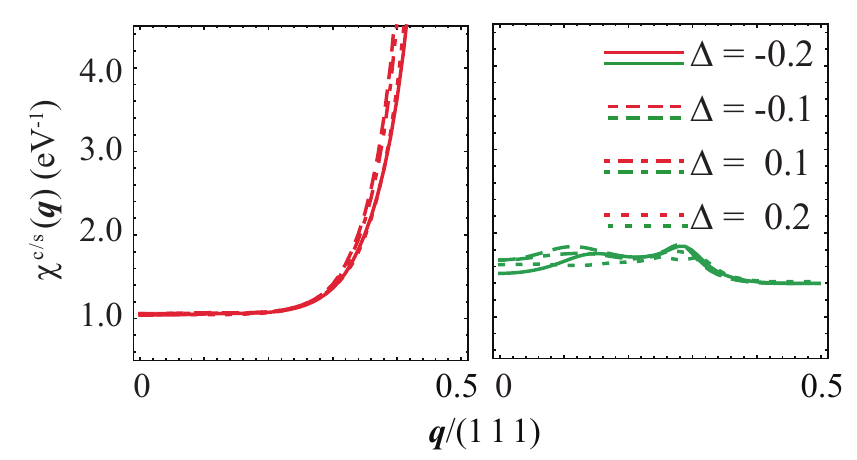}
  \caption{\label{Sfig:cf} Crystal field dependence of $\chi^c(\bm{q})$ (red) and $\chi^s(\bm{q})$ (green) for NdNiO$_3$ at 300 K.}
\end{figure*}

The crystal field $\Delta$ is included in our calculation by adding an energy difference to the onsite terms of the bulk Hamiltonian, i.e. we set $\epsilon'_{x^2-y^2}=\epsilon_{x^2-y^2}+\frac{1}{2}\Delta$ and $\epsilon'_{3z^2-1}=\epsilon_{3z^2-1}-\frac{1}{2}\Delta$. Therefore, positive (negative) $\Delta$ means more populated $x^2-y^2$ ($3z^2-1$) orbital. Figure~\ref{Sfig:cf} shows the results for $\Delta$ = -0.2, -0.1, 0.1, and 0.2 eV, corresponds to an orbital polarization [S3] of 17.5\%, 7.9\%, -10.0\%, and -18.8\%, respectively. Both $\chi^c(\bm{q})$ and $\chi^s(\bm{q})$ remains largely unaffected. 

We note that in reality, the crystal field induced by the substrate strain is usually accompanied by a change of the bandwidth, which we do not consider here. 

\section{Response functions in heterostructures}

The heterostructures can be viewed as finite system in the growth direction. For a $N_L$-layer slab (with each layer at $r_z$ = 1, $\dots$, $N_L$), the real-space fermionic operators can be defined as
\begin{equation}
  c^\dag_{\bm{r}} = \sqrt{\frac{2}{N_{xy}(N_L+1)}}\sum_{\bm{k}}c^\dag_{\bm{k}_{xy}k_z} e^{-\mi \bm{k}_{xy} \bm{r}_{xy}}\sin{k_z r_z}
\end{equation}
where we have separated the infinite in-plane ($xy$) and finite ($z$) parts, and $\bm{r}=\{\bm{r}_{xy},r_z\}$ ($\bm{k}=\{\bm{k}_{xy},r_z\}$). The orbital indices are left out here for brevity. The $k_z$ has only discrete values $k_z = l \pi/(N_L+1)$ with $l$ integers from 1 to $N_L$. For $r_z \notin [1,N_L]$, it is easy to check that $c^{(\dag)}_{\bm{r}}=0$, i.e. the wave-function vanishes outside of the slab. The total density $n=1/N\sum_{\bm{r}} c^\dag_{\bm{r}}c_{\bm{r}} = 1/N\sum_{\bm{k}} c^\dag_{\bm{k}}c_{\bm{k}}= 1/N_L\sum_{k_z}(1/N_{xy}\sum_{\bm{k}_{xy}}c^\dag_{\bm{k}_{xy},k_z}c_{\bm{k}_{xy},k_z})$, where $N=N_{xy}N_L$. This means for the finite system, its total occupation can be calculated by integrating the bulk density of states at different $k_z$ values. Similarly, the response function 
%\begin{equation}
%\begin{split}
%  \left< n(\bm{q}) n(\bm{-q})\right> & = \left( \frac{2}{N_{xy}(N_L+1)} \right)^2  \left< \sum_{\bm{k},\bm{k}',\bm{k}'',\bm{k}'''}\sum_{\bm{r},\bm{r}'} c^\dag_{\bm{k}}c_{\bm{k}'}c^\dag_{\bm{k}''}c_{\bm{k}'''} e^{\mi (\bm{k}_{xy}-\bm{k}'_{xy}+\bm{q}_{xy})\bm{r}_{xy}} e^{\mi (\bm{k}''_{xy}-\bm{k}'''_{xy}-\bm{q}_{xy})\bm{r}_{xy}'} \right. \\
%     & \phantom{\left( \frac{2}{N_{xy}(N_L+1)} \right)^2} \quad \quad \left. \sin{(k_z r_z)} \sin{(k_z' r_z)} \sin{(k_z'' r_z')} \sin{(k_z''' r_z') e^{\mi q_z (r_z-r_z')}} \vphantom{\sum_{\bm{r},\bm{r}'}} \right> \\ 
%  & = \left( \frac{2}{N_L+1} \right)^2 \left< \sum_{\bm{k}_{xy},\bm{k}_{xy}''} \sum_{k_z,k_z'} c^\dag_{\bm{k}_{xy},k_z}c_{\bm{k}_{xy}+\bm{q}_{xy},k_z'}c^\dag_{\bm{k}_{xy}'',k_z'}c_{\bm{k}_{xy}''-\bm{q}_{xy},k_z} \right. \\
%  & \phantom{\left( \frac{2}{N_L+1}\right)} \left.  \right>,
%\end{split}
%\end{equation}
%and subsequently the response function
\begin{equation}
  \chi^0(\bm{q},\mi \Omega_n) = -\frac{1}{\beta N_L} \sum_{\bm{k}_{xy}, k_z, k_z',  m} \alpha_{k_z,k_z',q_z} G^0(\bm{k}_{xy},k_z,\mi \omega_m) G^0(\bm{k}_{xy}+\bm{q}_{xy},k_z',\mi (\omega_m+\Omega_n))
\end{equation}
with $\alpha_{k_z,k_z',q_z} = \left[(\sum_{r_z} \sin{k_z r_z}\sin{k_z' r_z}\cos{q_z r_z})^2 + (\sum_{r_z} \sin{k_z r_z}\sin{k_z' r_z}\sin{q_z r_z})^2\right]$.

\begin{figure}[!htb]
    \centering
    \begin{minipage}[b]{0.54\textwidth}
        \centering
        \includegraphics[width=0.95\linewidth]{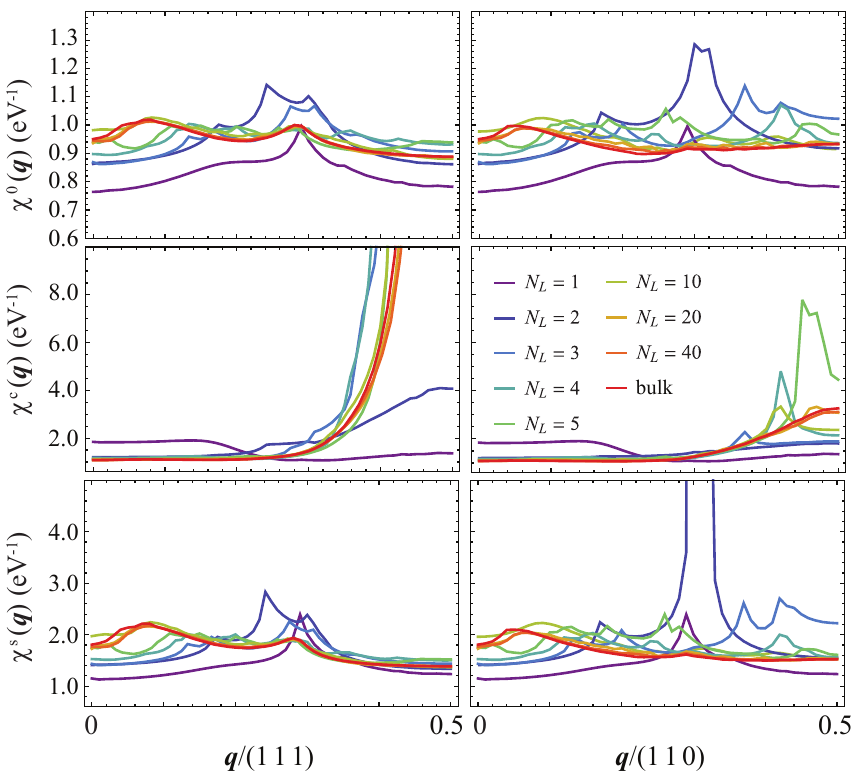}
        \caption{\label{Sfig:chislab} Thickness dependence of $\chi^0(\bm{q})$, $\chi^c(\bm{q})$, and $\chi^s(\bm{q})$ along (111) and (110) at 10~K.}
    \end{minipage}%
    \hfill
    \begin{minipage}[b]{0.44\textwidth}
        \centering
        \includegraphics[width=0.9\linewidth]{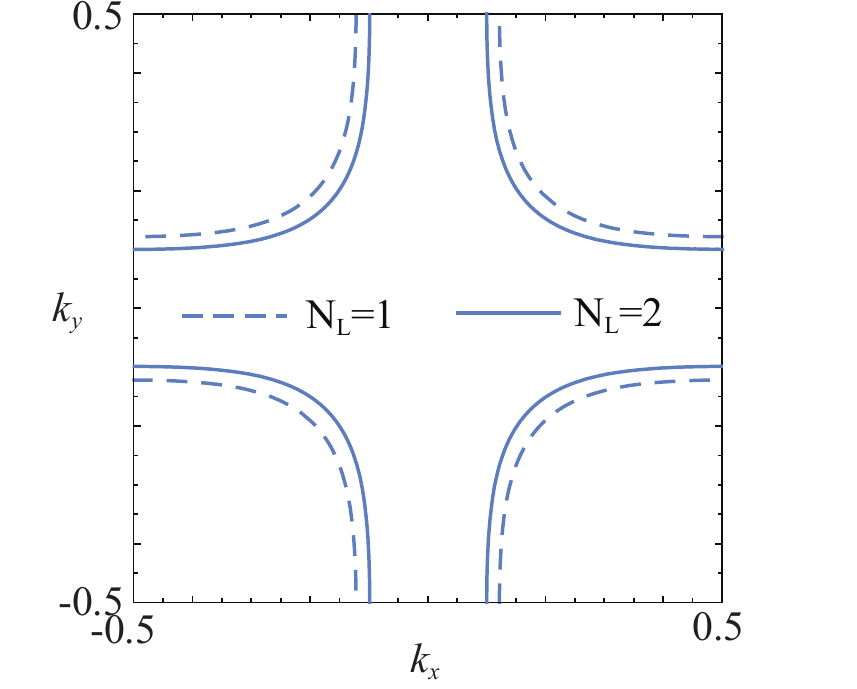}
        \caption{\label{Sfig:FSslab} Fermi surfaces for $N_L=1$ (dashed, $k_z=\pi/2$) and $N_L=2$ (solid, $k_z=\pi/3$ or $2\pi/3$).}
    \end{minipage}
\end{figure}

Figure~\ref{Sfig:chislab} shows the calculated $\chi^0(\bm{q})$, $\chi^c(\bm{q})$, and $\chi^s(\bm{q})$ for NdNiO$_3$ with different thicknesses. Considering the dimensional crossover, both $q$ paths along(111) and (110) are plotted. Note that for $N_L=1$, $q_z$ is not relevant anymore, which is evident from the fact that $\chi^{0/c/s}(\bm{q})$ are identical along the two plotted paths. A general trend of suppressed charge instability can be observed with decreasing $N_L$, and the divergence is removed for $N_L\leq2$. The behavior of $\chi^s(\bm{q})$ is more oscillatory with changing $N_L$, as it is originated from intra-orbital nesting and thus more sensitive to the detailed Fermi surface shape. For consistency we plotted always the maxima of $\chi^s(\bm{q})$ in Fig.~\ref{fig:confine} around $\bm{q}_s$, although for some cases a slightly larger value is found close to $(000)$. Note that this does not change the result qualitatively. For $N_L=2$, optimal nesting conditions are realized (Fig.~\ref{Sfig:FSslab}), with large flat sections parallel to each other.

\section{Constrained DFT+$U$ calculations}

\begin{figure*}[tb]
  \includegraphics[width=0.5\textwidth]{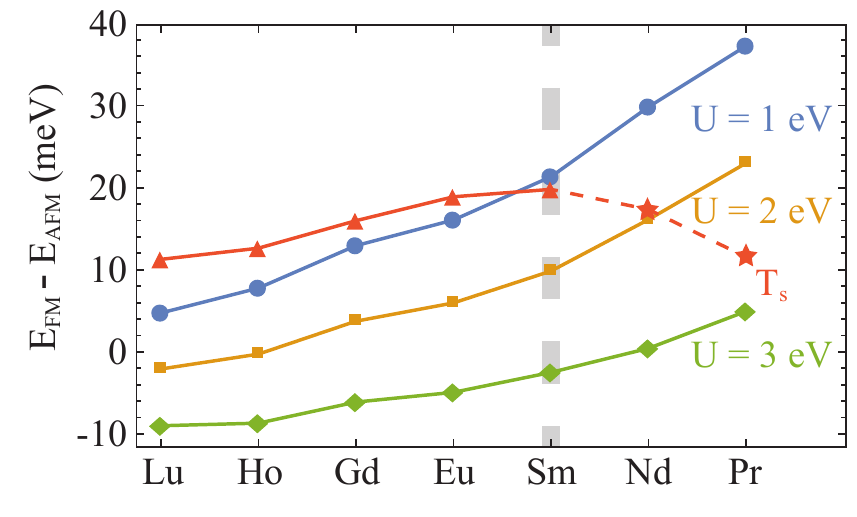}
  \caption{\label{Sfig:Eafm} Energy difference between the AFM and FM configuration for $U=$ 1, 2, and 3~eV, in comparison to the experimental $T_s$ (in eV).}
\end{figure*}

To estimate the material dependent tendency towards the antiferromagnetic order for the bulk \RNO{} in the monoclinic $P2_1/n$ phase, we performed DFT+$U$ with the VASP (Vienna ab initio simulation package) code [S4] using the generalized gradient approximation GGA-PBE functional [S5]. To capture the magnetic order with $\bm{q}_s$, we define a supercell with lattice vectors $\bm{a}'=\bm{b}$, $\bm{b}'=-\bm{a}+\bm{c}$, and $\bm{c}'=\bm{a}+\bm{c}$, where $\bm{a},\bm{b},\bm{c}$ are the lattice vectors for the original $P2_1/n$ unit cell. The magnetic ordering vector is then $\bm{q}'_s=(001)$ in the new definition. By applying $U\geq$ 1 eV, an antiferromagnetic (AFM) state in agreement with the experimental wave vector is stabilized. To obtain the relative energy gain of the AFM state for each material, we further performed calculations with constrained ferromagnetic (FM) configuration. The energy differences (per Ni) between the two magnetic states calculated with different $U$ values are plotted in Fig.~\ref{Sfig:Eafm} together the experimental ordering temperature $T_s$ (converted to energy). Although the stability of the AFM phase depends on the value of $U$, the relative energy gain follows the same material trend regardless of different $U$ values and is in agreement with that of the experimental $T_s$ from $R=$ Lu to Sm. The discrepancy for $R=$ Nd and Pr highlights the importance bond order for the AFM transition (see main text).

%The superexchange can then be calculated as $J_{ex}=(E_{AFM}-E_{FM})/6$, where $E_{AFM}$ ($E_{FM}$) is the energy per Ni site for AFM (FM) state, and 6 is the number of flipped next-nearest-neighbor spin pairs for each Ni between the two magnetic configurations.

\noindent\makebox[\linewidth]{\resizebox{0.3333\linewidth}{2pt}{$\bullet$}}\bigskip

[S1] Z. Zhong and P. Hansmann, arXiv:1611.08689 (2016).

[S2] J. Hoffman et al., \emph{Phys. Rev. B} \textbf{88}, 144411 (2013).

[S3] M. Wu et al., \emph{Phys. Rev. B} \textbf{88}, 125124 (2013).

[S4] G. Kresse and D. Joubert, \emph{Phys. Rev. B} \textbf{59}, 1758 (1999).

[S5] J. P. Perdew, K. Burke, and M. Ernzerhof, \emph{Phys. Rev. Lett.} \textbf{77}, 3865 (1996).

\end{document}